Original Paper

# Development and User Experiences of a Novel Virtual Reality Task for Poststroke Visuospatial Neglect: Exploratory Case Study

Andrew Danso[1], PhD; Patti Nijhuis[1], PhD; Alessandro Ansani[1], PhD; Martin Hartmann[1], PhD; Gulnara Minkkinen[2], MA; Geoff Luck[1], PhD; Joshua S Bamford[1], PhD; Sarah Faber[3,4], PhD; Kat R Agres[5], PhD; Solange Glasser[6], PhD; Teppo Särkämö[1,7], PhD; Rebekah Rousi[8], PhD; Marc R Thompson[1], PhD

[1]Centre of Excellence in Music, Mind, Body and Brain, University of Jyväskylä and University of Helsinki, Jyväskylä and Helsinki, Finland
[2]Department of Music, University of Jyväskylä, Jyväskylä, Finland
[3]Simon Fraser University, Burnaby, Canada
[4]University of Toronto, Toronto, ON, Canada
[5]Centre for Music and Health, Yong Siew Toh Conservatory of Music, National University of Singapore, Singapore, Singapore
[6]Melbourne Conservatorium of Music, The University of Melbourne, Melbourne, Australia
[7]Cognitive Brain Research Unit, Department of Psychology, Faculty of Medicine, University of Helsinki, Helsinki, Finland
[8]School of Marketing and Communication, Communication Studies, University of Vaasa, Vaasa, Finland

**Corresponding Author:**

Andrew Danso, PhD
Centre of Excellence in Music, Mind, Body and Brain
University of Jyväskylä and University of Helsinki
Seminaarinkatu 15, Jyväskylän yliopisto
Jyväskylä and Helsinki PO Box 35, FI-40014
Finland
Phone: 358 406643034
Email: andrew.a.dansoadu@jyu.fi

## Abstract

**Background:** Visuospatial neglect (VSN) affects spatial awareness, leading to functional and motor challenges. This case study explores virtual reality (VR) as a potential complementary tool for VSN rehabilitation.

**Objective:** Specifically, we aim to explore the initial experiences of patients and physiotherapists engaging with a novel protocol, using an audiovisual cue task to support VSN rehabilitation.

**Methods:** A preliminary VR task integrating audiovisual cues was co-designed with 2 physiotherapists. The task was then tested with 2 patients with VSN over 12 sessions. The intervention focused on engaging neglected spatial areas, with physiotherapists adapting the task to individual needs and monitoring responses.

**Results:** Initial testing with 2 trainee physiotherapists indicated high usability, engagement, and perceived safety. Two patients with VSN completed 12 VR sessions. For Patient A, completion times increased following the introduction of an audio cue, though modeling indicated a nonsignificant linear trend (β=0.08; *P*=.33) and a marginally significant downward curvature (β=–0.001; *P*=.08). In contrast, Patient B showed a significant linear decrease in completion times (β=–0.53; *P*=.009), with a quadratic trend indicating a performance minimum around session 10 (β=0.007; *P*=.04). Intraweek variability also decreased. Motor scores (Box and Block Test and 9-Hole Peg Test) remained stable, and subjective feedback indicated improved mobility confidence and positive task engagement.

**Conclusions:** Further research with larger cohorts is needed to confirm the VR task's utility and refine the intervention.

# Introduction

## Background

Following a stroke, approximately 30% of stroke survivors experience neglect [1]. Neglect is a neurological disorder that poses significant challenges for rehabilitating behavioral deficits, including motor functions and perceptual-cognitive impairments, such as spatial awareness. Among the various types of neglect, visuospatial neglect (VSN) stands out as a specific subset. This is characterized by a failure to attend to objects or events within a defined region of the visual field, commonly affecting the left side [2]. These deficits increase the risk of falling and contribute to caregiver burden [3]. Conventional rehabilitative interventions typically involve manual interactions between patients and therapists, which can be physically demanding for practitioners and patients, leading to disempowerment, boredom, and reduced motivation when activities lack autonomy or engagement [4]. The integration of technology-based modalities, such as serious games and virtual reality (VR), with conventional rehabilitative interventions has emerged as a promising approach to engage patients with poststroke neglect [5]. When used in conjunction with traditional rehabilitation, including physiotherapy interventions, these modalities offer the potential for a more motivating treatment experience. Despite the potential benefits of these approaches, the use of audiovisual cues within VR adjunctively with physiotherapy remains poorly understood in terms of the subjective experience of patients with VSN. Consequently, the objectives of this case study are (1) to develop a real-time VR-based physiotherapy training solution tailored for individuals with VSN and (2) explore how audiovisual cues may influence the performance and rehabilitation experience of physiotherapists and individuals living with VSN during interaction with the VR-based training solution across 12 sessions.

## Visuospatial Neglect

VSN, a common cognitive deficit following a stroke, is characterized by persistent spatial inattention, often manifesting unilaterally [2,6]. Patients with VSN struggle to acknowledge or respond to visual stimuli presented on the side opposite to the damaged hemisphere, often behaving as if that side of their visual field does not exist [7,8].

VSN is typically associated with damage to the posterior-parietal cortex of the right hemisphere. However, recent lesion mapping studies suggest a high degree of variability regarding the anatomical basis for neglect [9] with the temporo-parietal cortex, frontal cortex [10], occipital cortex [11], cerebellum [12-14], and even subcortical regions [15] have been linked to neglect. Furthermore, it has been associated with disconnections in white matter tracts, such as the superior longitudinal, inferior longitudinal, and inferior fronto-occipital fasciculi [9]. These varied findings highlight the complexity of VSN's neuroanatomical correlates.

Neglect mainly affects higher-level spatial processing modalities, such as visual and auditory spatial processing [16]. However, the empirical relationship between visual and auditory tasks with neglect remains unclear [8]. As studies have simulated multisensory (typically audiovisual and tactile) training procedures, improvements have been observed after training that used temporally congruent audiovisual input [16,17]. Therefore, the exploration of multisensory and specifically audiovisual training procedures is warranted.

## Physiotherapy and Patient-Centered Treatment

Poststroke rehabilitation programs commonly feature physiotherapy to address motor and sensory impairments [18]. Physiotherapy is a vital primary care service within formal health care systems, aiming to sustain optimal physical functioning through various nonpharmacological interventions, such as progressive exercises [19,20]. Previous research [21,22] indicates that task-specific repetitive practice is essential for attaining lasting improvements in motor learning and motor function.

Recent trends in physiotherapy further emphasize the importance of patient-centered treatment [23]. Patient-centric physiotherapy treatment involves physiotherapists providing support to empower patients by providing emotional and physical assistance, alleviating fears and anxiety, and involving family and friends in treatment and care plans when possible. Patient-centered support often takes various forms of communication, including both verbal and nonverbal methods, such as tactile interactions and patient education [24]. Given the multifaceted nature of conditions, such as VSN, where multiple aspects of perception and movement may be affected, patient-centered support is particularly pertinent, as it addresses a wide range of dimensions in the recovery process.

Since the reaching and grasping skills of these patients are often limited, physiotherapy programs targeted for poststroke neglect rehabilitation include grasping training. Grasping training for poststroke neglect aims to improve spatial representation ability, as well as an enhancement in reaching, interacting, and grasping skills toward the neglected area or environment [25]. This often takes the form of congruent visual scanning training and motor rehabilitation tasks [26,27]. Studies suggest grasping training using methods such as home-based programs and custom-developed VR simulations is beneficial for patients with poststroke neglect to develop reaching and grasping skills that can be tailored to individual needs [28], which improves their ability to grasp objects [29-31].

## VR and Neglect Rehabilitation

VR has emerged as a promising technology to be used adjunctly with physiotherapy, aiming to influence physical behaviors and movements within immersive, computer-generated environments. Sensory-motor tasks in VR offer several distinct advantages for physiotherapists. VR provides a safe setting for patients to engage in realistic and repetitive movements, either as an adjunct to conventional physiotherapy or in tandem with it, under the real-time supervision of therapists [5]. Evidence from various studies suggests that VR can improve the frequency of motor tasks in poststroke rehabilitation by increasing practice intensity [32], improving hand function [33], and promoting neuroplastic changes [34].

Recent studies have demonstrated VR's effectiveness in various stages of VSN management, ranging from diagnosis [35] and assessment [36] to motivation and rehabilitation [37]. This is in part due to VR's capacity to create immersive and controllable training environments, enhancing patient engagement and motivation, potentially leading to better treatment adherence and outcomes [38]. For rehabilitation, several studies have used VR to simulate realistic grasping training through hand grasp motions, showing promising results [39,40]. The engaging, adaptable, and measurable aspects of VR thus prove it to be a promising tool for VSN rehabilitation.

### VR and Audio-Tactile Cueing in Neglect Rehabilitation

In VSN rehabilitation, audio-tactile cues enhance the immersive effects of VR by directing attention toward the neglected space through multimodal sensory engagement, thereby promoting orientation and visual awareness on the affected side. Studies by Knobel et al [41] and Leitner and Hawelka [42] provide evidence that audio-tactile cueing in VR settings with patients with VSN can effectively improve patients' attentional orientation and head movement toward stimuli, assisting them to overcome rightward orientation biases. VR interventions can provide a structured and repeatable therapeutic experience, aligning with neuropsychological approaches (eg, prism adaptation therapy, a rehabilitation technique involving the use of prism glasses to shift the visual field and correct for visual displacement). Prism adaptation therapy has been integrated into VR environments, leading to more effective rehabilitation outcomes [43] as well as visual scanning training [42,44]. Phasic alertness (the brief adaptive increase in arousal that occurs in anticipation of an upcoming warning stimulus, see eg given by Posner [45]) has also been shown to be positively influenced with audiovisual cueing, leading to improvements in the balance of visual attention in patients with neglect [46]. Auditory cues can trigger fast, automatic shifts in spatial attention, suggesting preservation of strong links between auditory and visual attention mechanisms in patients with neglect. Sustained long-term improvements have been found following intensive and prolonged multisensory audiovisual stimulation [47].

This case study addresses a critical gap in the co-design and iterative development of VR-based interventions tailored specifically for hand grasping training in patients with VSN. Unlike generalized VR applications in rehabilitation, this intervention was designed through interdisciplinary collaboration with physiotherapists to integrate audiovisual cueing within a hand-grasping task, offering a novel approach to VSN rehabilitation [48]. For instance, existing VR-based interventions [49] have primarily focused on perceptual training through visual scanning tasks or general attentional cueing, whereas the system developed here aims to incorporate elements of compensatory motor initiation, less commonly addressed in this context. Compensatory motor initiation refers to the use of alternative motor strategies, such as gaze shifts, to facilitate movement toward the neglected hemispace, particularly in patients with VSN who exhibit impaired initiation on the contralesional side [50]. Accordingly, this VR intervention distinctly explores the integration of physiotherapist-informed design components, such as adjustable audiovisual cueing and targeted hand-grasping tasks, to address compensatory motor initiation and spatial attention. The structured co-design process included iterative testing and refinement to align the intervention with patient-specific needs and therapeutic goals [51]. In addition, exploring how such tasks influence individual patient experiences over multiple sessions provides valuable insights for personalizing rehabilitation strategies, addressing a critical need for evidence in this domain [1,52]. Furthermore, understanding how such a task influences the experience of individual patients over a series of physiotherapy sessions is unknown. Therefore, the following research question directed the study:

RQ. What are the initial experiences of patients and physiotherapists using a custom-developed VR-based hand grasping training protocol?

Accordingly, the aims of this study were 2-fold: (1) to develop a solution using audiovisual cueing to be used during real-time physiotherapy training and (2) explore the initial experiences and perceptions of patients and physiotherapists regarding the use of audiovisual cueing in the VR task.

## Methods

### Intervention Development

The case study presents a VR-based physiotherapy intervention designed for hand grasping training in the rehabilitation of VSN. This intervention uses VR to produce customizable visual and audio cues in its environment, aiming to address the requirements of individual patients with VSN. For instance, elements such as the timing, location, and dynamics of these cues can be adjusted to optimize the patient's training experience. In this intervention, participants engage in a VR task. The VR task is a single trial where a ball, serving as the visual cue, appears to the left within the VR environment, preceded by an audio cue to signal its location. The ball bounces in a fixed vertical up-down trajectory. This allowed users to plan their motor responses, such as grasping. Participants were tasked with grasping the ball as quickly as possible and were limited to grasping 1 ball per trial. In this study, we first focused on usability, followed by user testing. Audio cues were introduced in the seventh week to streamline the complexity of the VR task. This procedure-based approach to VR intervention design involves a structured process of usability testing and the phased introduction of multisensory cues to enhance task performance, aligning with principles of effective VR rehabilitation training [48].

Aligned with the UK Medical Research Council's (MRC) guidelines for developing complex interventions [51,53], this case study emphasizes a structured and evidence-based approach to intervention development. The UK MRC's emphasis on exploring feasibility and acceptability in the early stages of intervention development was addressed through detailed user testing (by both physiotherapists and patients with VSN) and design, to refine the VR task and ensure it met necessary user requirements.

## Design and Implementation

For the purposes of this case study, an experimental VR environment was developed in Unity3D (Unity Software Inc). As shown in Figure 1, the VR environment contained multiple objects. Using Unity3D, the software developer crafted an experimental VR environment that featured a prominent visual cue in the form of a red ball. In addition, an auditory directional cue was used. The audio cue directed the user to the ball's location, and the task was considered accomplished upon successfully grasping the ball.

**Figure 1.** First-person perspective of the virtual reality task environment.

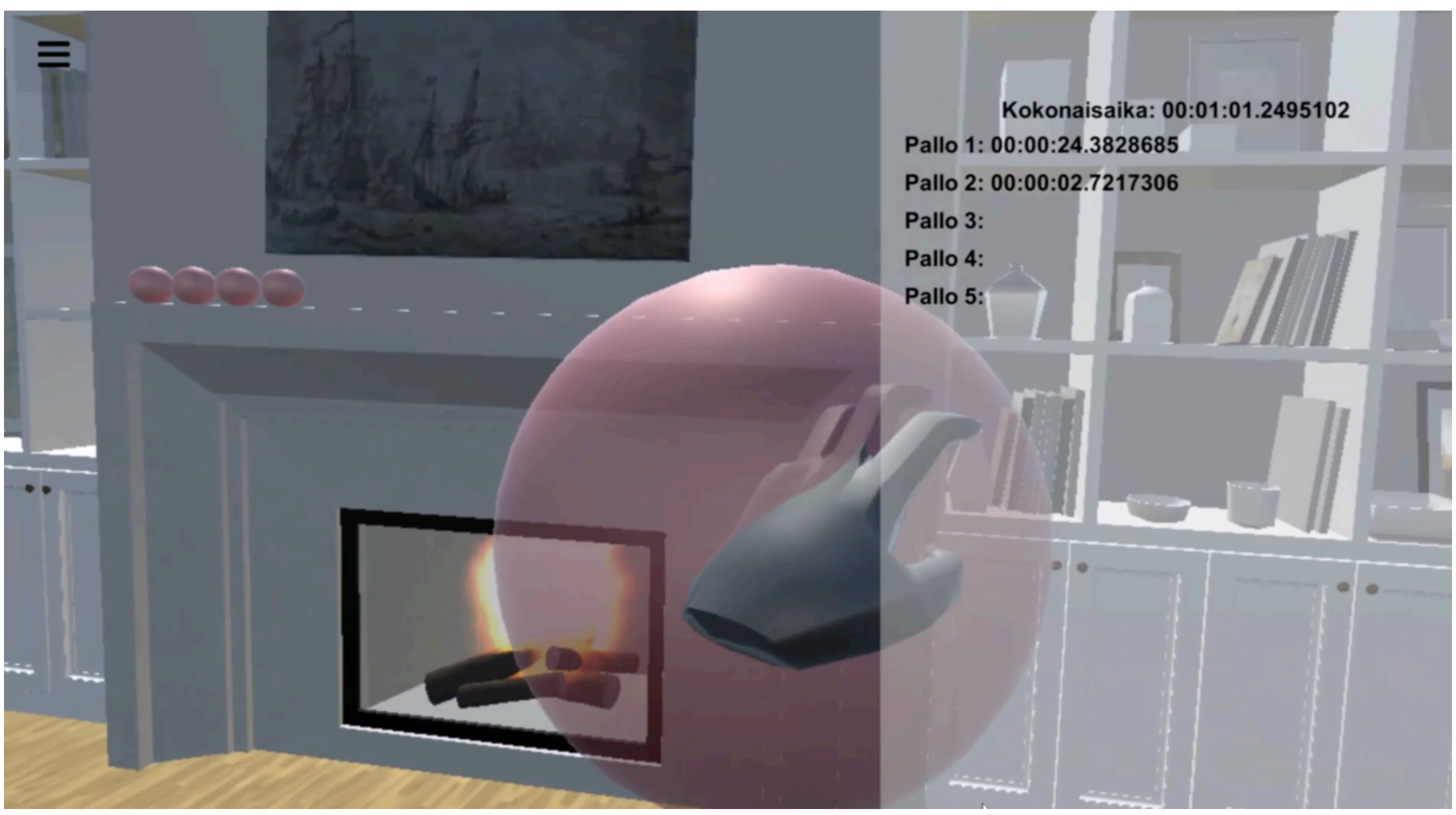

## Directional Cue Design

### Visual Cue

Figure 1 illustrates a red ball designated as the focal point for participant interaction, serving as a visual cue to direct their attention toward the task of grasping during trial runs. The figure depicts a VR task environment as viewed from a first-person perspective. In the center of the room, a red ball serves as the primary visual cue. The user's left hand is shown reaching for the red ball, indicating the action-based element of the task. The ball appears in the center, as this was captured during early development and captured only for example purposes. The interface also includes a timer display with labels in the Finnish language, such as "kokonaisaika," meaning "total time," and "Pallo 1" through "Pallo 5," meaning "ball 1" through "ball 5," respectively. These labels are followed by time stamps, revealing the duration taken to interact with each ball; for instance, "Pallo 1" took 24.38 seconds, and "Pallo 2" took 2.72 seconds. The sequence suggests that the user will engage with a series of 5 such balls throughout the exercise. The surrounding environment is minimally designed with a neutral color palette, emphasizing focus on the task elements. The Unity 3D rendered scene of the VR environment is presented in Figure S1 of the Multimedia Appendix 1

Developed using Unity 3D, the designated grasping zone is active from the ball's periphery to its center and is detected by Unity's collision system. The ball is designed to move vertically within the space (ie, vertical cues have been found to influence spatial orientation and potentially aid in rehabilitation of patients with VSN, see Lafitte et al [54]). The dynamics of this movement, such as the bounce speed, can be adjusted from 1 millisecond to 60 seconds through the application settings.

For this cue, a range of available adjustments was deemed important, as Golay [55] suggests that the effectiveness of cues in neglect rehabilitation can vary depending on the interval between the cue and the target. The starting position of the red ball correlates with the user's spatial location in the VR environment, which is determined by the positioning of their head-mounted display (HMD). The ball ceases movement when the participant's hand is near, simulating interaction. Successful grasping is indicated by the ball's disappearance. The appearance of the ball is designed to occur within the participant's left visual field (based on the [56] reach task consistently using targets appearing outside a central fixation point), determined by the spatial audio cues' effect on the participant's orientation in the HMD. The red ball appears approximately 61 centimeters (2 feet) from the participant, facilitating reach and interaction (eg, [57]).

## Audio Cue

The audio cue was made by using the spatial sound capabilities of Unity3D 5.3, with the spatial blend parameter set to full 3D, allowing for precise auditory localization in conjunction with visual elements. Unity's spatial audio geometrically simulates sound sources within the environment, with the auditory cues emanating from the expected ball appearance location relative to the user's HMD position, facilitated by a head-related transfer function (HRTF) system. HRTF technology mimics how sound is affected by the listener's head and ears, providing a naturalistic sound perception based on directionality. The audio cue lasted 2.61 seconds and served to alert users to the specific location where visual stimuli would appear. This was based on prior research by Yoshizawa et al [58] demonstrating that a cue lasting 2-3 seconds effectively directed attention toward the neglected side during VR rehabilitation tasks for patients with hemispatial neglect. Studies by Dozio et al [59] and Knobel et al [41] suggest that short-duration audio cues are both beneficial and suitable in VR interventions for VSN rehabilitation. An interstimulus interval of 105 milliseconds between the auditory and subsequent visual cue was optimized to prepare the patients for grasping the red ball (visual cue). The frequency spectrum of the auditory cue showed a prominent peak at approximately 500 Hz (≈−30 dB) within the 50–20,000 Hz range, with additional smaller peaks across the low-frequency range. Figure 2 illustrates the frequency spectrum of the auditory cue. Figure 3 provides a 3D visualization of the auditory cue.

**Figure 2.** Frequency spectrum plot of auditory cue.

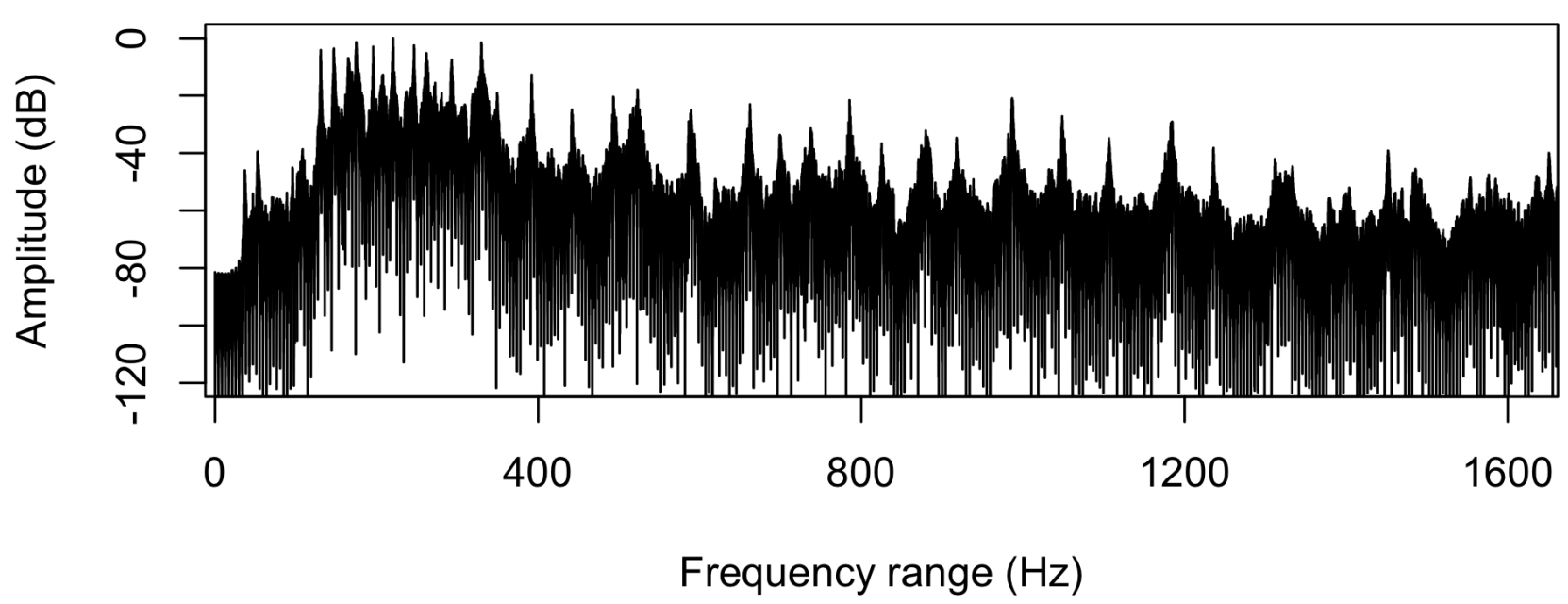

**Figure 3.** 3D visualization of auditory cue. The x-axis—Time (s)—denotes time in seconds, the y-axis—Channel—displays the left and right audio channels, and the z-axis—Amplitude (k)—illustrates the amplitude scaled in kilounits. The color gradient, as indicated by the color bar labeled "Amplitude Variation," visually depicts the amplitude fluctuations within the auditory cue.

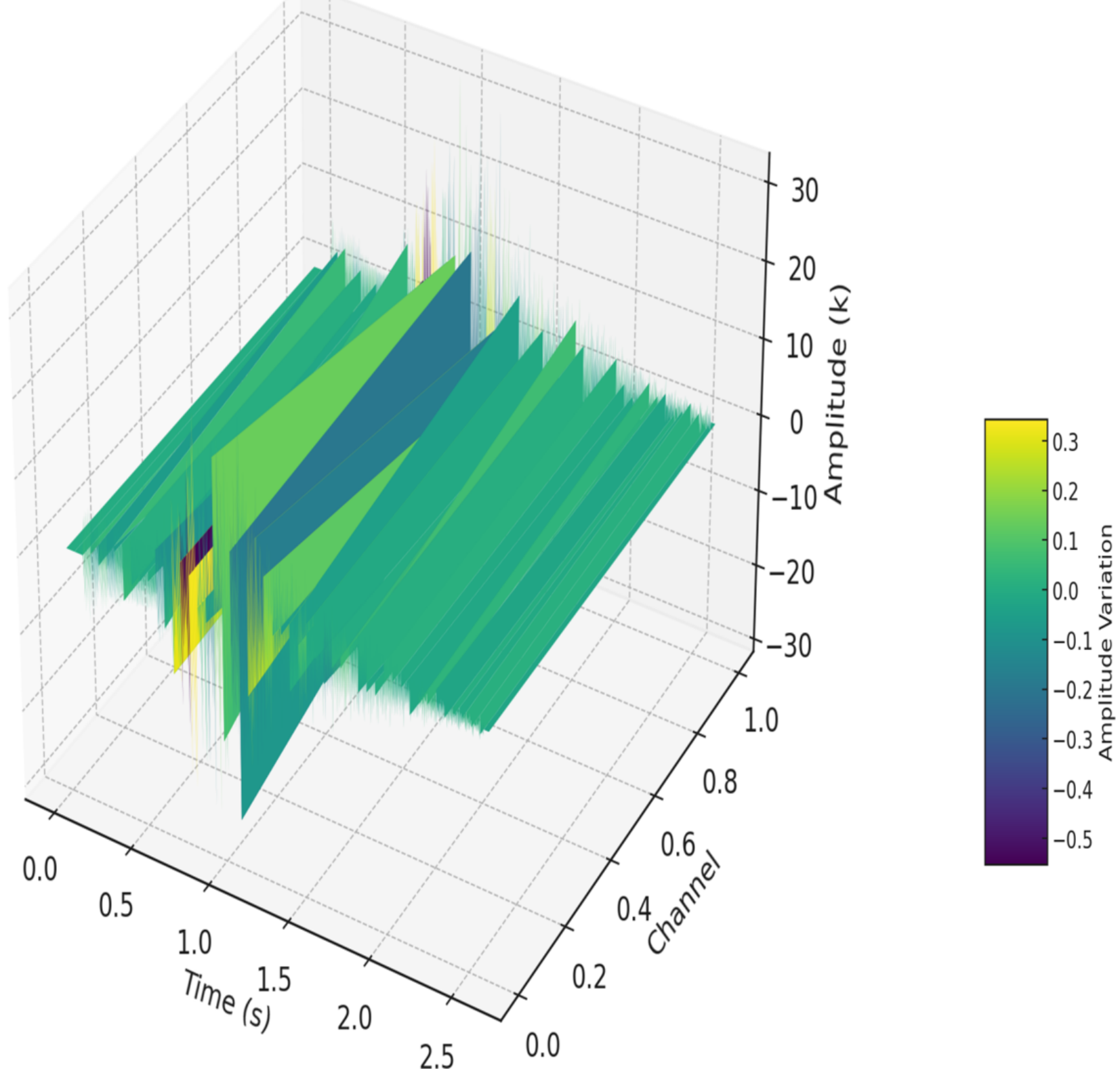

## Task Parameters and Environment Design

### VR Task Description

The VR task includes 1 trial with a ball appearing to the left (15° to the left), within a 30° horizontal plane and a 50° vertical plane within the limits of the VR room. To complete each trial, participants are required to successfully grasp the floating ball as fast as possible (includes a 5-minute timeout period for managing patient fatigue, [22,23,60]). Upon appearing, the ball bounces with a vertical up-down trajectory in the room and stays bouncing within a fixed vertical trajectory until the participant grasps the ball. Participants were limited to grasping 1 ball once per trial. To initiate the appearance of a new ball, participants were required to rotate their trunk and direct their gaze toward the center point of the field of view (FOV). This central gaze point was represented in the VR software as a painting positioned above a fireplace object. The process was marked by a countdown timer, starting from 3 and concluding at 0, at which point a new ball was generated. The duration of each trial was measured (in ms) until a successful grasp occurred. The ball's visual stimuli were depicted through its appearance in the VR environment. An audio cue of where the ball will appear across the 30° horizontal plane was activated prior to the ball appearing to alert the patient to the appearance location (interstimulus interval between the audio cue and visual cue=105 ms, ie, [41]). There were 15 trials in each location across the 30° x 50° degree plane. The timer's data aimed to provide insights into how target grasping efficiency is affected by the size and distance of objects. In addition, multiple trials were incorporated and the elimination of manual restarts to improve the interaction process and reduce the cognitive load associated with initiating new trials. Data output was presented in the form of a text file, which recorded participant response times for each trial, audio cue sounds, and trial dates.

### Task Development Aims

The development of this VR task is specifically tailored for inclusion in poststroke physiotherapy rehabilitation, building upon the Reach Task conceptualized by Mattingley et al [56]. The original task involved participants reaching toward a stimulus at the edge of their visual field. Our adaptation for VR purposes follows this principle, focusing on the neglected visual field to help distinguish perceptual deficits from motor control difficulties.

To support physiotherapeutic goals, the task encourages the use of compensatory strategies—alternative motor patterns developed to adjust for lost function, as described by Levin et al [61]. Patients with VSN engaging in these motor patterns is critical for fostering attentional shifts [62]. Integrating audio cues with visual targets is designed to enhance anticipatory behavior [55], supporting patients with neglect to proactively direct their attention and gaze toward the task at hand.

### Applying Fitts Law as a Guiding Principle for Task Difficulty

Fitts law is a psychological principle stating that the difficulty of a perceptual-motor task, such as pointing or selecting targets, is a function of target size and distance [63]. The smaller the target size, the slower and more difficult it is for individuals to accurately reach or activate the target. This provided a framework for reducing the difficulty by enlarging the target and thereby increasing the accessibility of user interactions within the VR environment. As part of subsequent development tasks, the software developer adjusted the FOV to 30° on the horizontal plane and 50° on the vertical plane using Unity's built-in parameters. This adjustment was chosen to reduce the distance to targets, thereby making the task more accessible for patients with VSN to successfully grasp objects (this also reduced the potential for cybersickness effects, eg, [60]). The adjustment of the FOV to specific angles on the horizontal and vertical planes is also made in relation to the effective size and position of the targets (ie, the balls) within the VR environment (see Figure 4 depicting a conceptual diagram representing the appearance location for the primary visual cue within the updated FOV parameters). The ball was set to appear at ground level, ascend vertically to the ceiling, and descend vertically back to its initial point of origin on the ground (up-down trajectory).

**Figure 4.** Illustrative figure depicting the field of view in the virtual reality task. This figure depicts the field of view in the virtual reality task, with 4 red spheres indicating potential appearance locations of the primary visual cue within a 15° range to the left of the central gaze. The black line represents the participant's position at coordinates x=3, y=1. “X=30°” and “YZ=50°” denote the maximum horizontal and vertical area visible to the participant. The 2 blue arrows indicate the breadth and height of the participant's potential visual engagement area during the activity. The diagram is conceptual and not drawn to scale; axis measurements of 7x7 meters (X and Y) and 3.5 meters (Z) are for reference only.

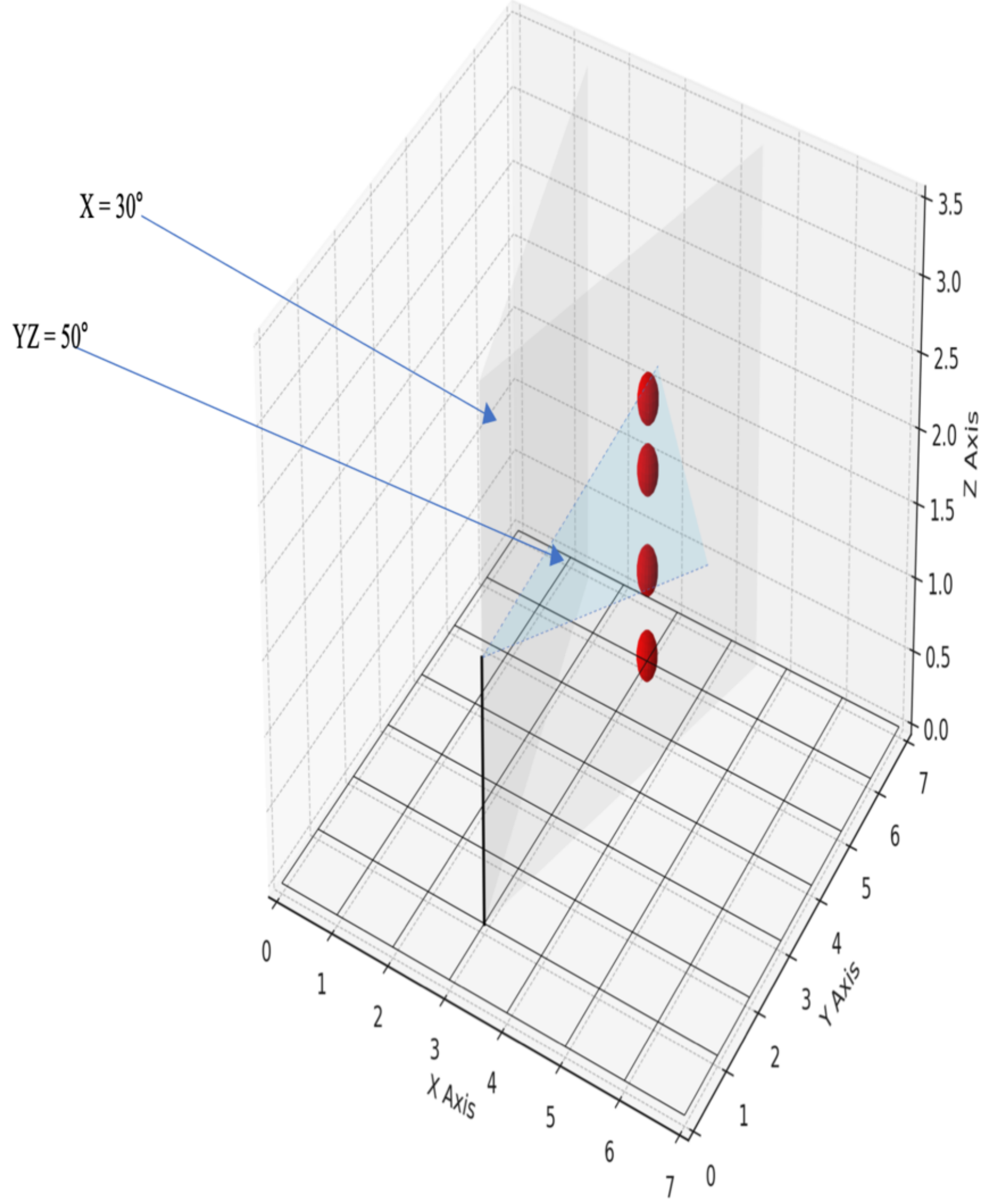

## High Tech Computer Corporation Vive Head Mounted Display

The VR intervention was delivered using the High Tech Computer Corporation (HTC) Vive HMD, a tethered head-mounted display equipped with lighthouse tracking technology for accurate room-scale motion capture. Its compatibility with HRTF audio makes it suitable for delivering spatial audio cues, essential for the task design. A detailed specification of the VR hardware setup, including display resolution, connectivity, and audio components, is provided as supplementary information (Section A in Multimedia Appendix 1).

## Leap Motion Controller

Hand tracking in the VR task was achieved using the Leap Motion Controller, a touchless optical tracking device that allows real-time monitoring of hand and finger movements [64]. This interface enabled intuitive grasping interactions without the need for handheld controllers. Haptic feedback was not included, in line with previous studies highlighting the complexity it introduces in poststroke rehabilitation tasks [64,65]. A specification of the Leap Motion Controller and how it pertains to this study is provided as supplementary information (Section B in Multimedia Appendix 1).

# *Usability and Preliminary Testing*

## Preliminary User Testing

During the initial development phase, collaboration took place between a physiotherapist from the anonymous physiotherapy clinic, who played the role of a user tester, and a software developer from (anonymous organization). For testing purposes, they used a PC-based system along with a tethered HTC Vive HMD and a Leap Motion Controller. Initially, challenges emerged related to latency, particularly concerning the responsiveness of the Leap Motion Controller device to the grasping gesture. To overcome this issue, they decided to externally mount the Leap Motion Controller on the HMD to enhance the tracking of hand movements and gestures, serving as a trigger point for task completion. Mounting Leap Motion Controllers on a VR HMD has been found to enlarge the tracking area for hand tracking in VR software programs, thereby improving user experience with enhanced hand and gesture tracking [66].

### Usability Assessment

To assess usability as well as the physiotherapists' perception of the task and system, 4 questions from the Technology Acceptance Model questionnaire [67], 2 questions from the Engagement in In-Game Questionnaire [68] and 1 question from the Safety Attitudes Questionnaire [69] were adapted and used for a usability assessment. The questions were answered using a 5-point Likert scale, ranging from "fully disagree" to "fully agree" (see Table S1 in Multimedia Appendix 1).

### *Ethical Considerations*

The study was conducted in accordance with the Declaration of Helsinki [70] and a favorable ethical statement from the Pirkanmaa Ethics Committee (984/2021). Written informed consent was obtained from all participants. Personal data were processed under a Data Processing Agreement in compliance with the GDPR (EU, 2016/679). Data were pseudonymized, stored securely on password-protected servers, and accessible only to authorized researchers. No identifiable images or personal identifiers are included. Participants were reimbursed for travel expenses but received no other compensation.

## Results

### *Feasibility and Usability Feedback (Physiotherapists)*

Before conducting tests with participants with VSN, 2 trainee physiotherapists from an anonymous physiotherapy clinic participated in VR task test sessions to gain insights into their experiences with the technology. Subsequently, these test sessions were immediately followed by a survey where the 2 physiotherapists provided feedback regarding their subjective experiences with the VR task. The survey was administered using Webropol (Webropol Oy) software. Each trainee used the application for approximately 10-15 minutes before completing the survey. Table 1 aimed at assessing (1) the physiotherapists' perceived ease of use of the system, (2) the physiotherapists' engagement while using the system, and (3) their perception of the safety of the VR task. The questions for usability (Question 1–Question 4), task engagement (Question 5), and safety (Question 7) were rated highly by both trainee physiotherapists. The question associated with difficulties in learning the task (Question 6) was rated low.

**Table 1.** Questionnaire results from 2 trainee physiotherapists.

| Survey question | Physiotherapist 1 rating | Physiotherapist 2 rating |
|---|---|---|
| I think the system is easy to use | 4 | 4 |
| Learning to use the system is not a problem | 4 | 5 |
| I enjoyed using the system | 4 | 4 |
| I would like to use the system in the future if I had the opportunity | 4 | 4 |
| Learning to use this VR[a] task was easy | 4 | 4 |
| Was learning the task difficult? | 2 | 2 |
| I would feel safe using this as a patient | 4 | 4 |

[a]VR: virtual reality.

Following feedback from the physiotherapists during initial testing, session length was set to approximately 10-15 minutes. While fatigue was not directly measured, this duration was selected to support tolerability and aligns with findings recommending shorter VR sessions (10-30 min) to reduce fatigue in poststroke rehabilitation [71].

### *Patient Characteristics and Baseline Function*

Patients were recruited between February and March 2022 from inpatients with poststroke neglect referred to care at the anonymous hospital/clinic, Finland. Inclusion criteria required participants to have a right hemisphere stroke with diagnosed neglect, be right-handed, be medically stable, be without hearing impairments, cognitive deficits (eg, learning difficulties), or hemianopia (loss of vision in one-half of the visual field), and be aged 18 years or older. In addition, patients were assessed for their physical and cognitive ability to perform the audiovisual VR task by a physiotherapist team. Patients were excluded if motor or communication impairments, as determined by the physiotherapists, were severe enough to prevent task participation or understanding of instructions. Eligible inpatients received detailed study information, and participation was discussed.

Based on the results of the previous phase of testing, and on the expertise of the physiotherapy clinic, we decided to use the VR task as part of physiotherapy sessions with 2 patients with poststroke neglect for the next phase of exploration. To preserve procedural integrity and ensure personalized care, sessions were conducted by 2 licensed physiotherapists from an anonymous hospital/clinic. Patient A was a 46-year-old male with left-sided hemiparesis and VSN, 1 year poststroke (Barthel Index: 70/100). Patient B, a 37-year-old female with hemiplegia and VSN, was 4 years poststroke (Barthel Index: 95/100). Both were right-handed and met inclusion criteria (full clinical profiles and ADL scores are presented in Table S1 and Section D in Multimedia Appendix 1).

The inclusion of 2 patients with differing clinical profiles, a mild case (Patient A's mild hemiparesis) and a more severe case (Patient B's severe hemiplegia), was a deliberate methodological choice consistent with early-phase intervention research (eg, UK MRC guidelines [51,53]). This heterogeneity enabled an initial assessment of the VR system's usability across a spectrum of functional symptoms and rehabilitation timelines. Such purposive sampling is supported in the development of complex interventions, where the goal is to evaluate feasibility, individual responsiveness, and context-specific implementation [51,53,72]. In neurorehabilitation, evidence shows that early inclusion of diverse patient profiles enhances understanding of task usability, supports iterative design, and informs future personalization strategies [1,41]. Furthermore, diverse case inclusion enables deeper insight into patient-centered customization [17,52] (see eg, [73] where VR task parameters were adapted to patients with varying upper-limb impairments, improving usability, and elevating future patient adherence to the intervention).

## Description of Patient Test Sessions

Physiotherapy interactions were standardized across both patients to ensure procedural consistency while supporting individual needs. Both licensed physiotherapists underwent training in the VR task and applied identical task parameters (eg, 1.19 s ball bounce, consistent audio cue use), emphasizing procedural integrity and a patient-centered framework [74]. Patients received uniform instructions and completed preparatory sessions to familiarize themselves with the VR environment. During sessions, physiotherapists monitored performance in real time and provided feedback based on individual motor behavior, such as compensatory strategies (eg, trunk rotation or delayed reaching). Therapists manually initiated each trial using the in-task menu (see Figure S2 in Multimedia Appendix 1). A detailed description of session setup, training, and interaction procedures is included as supplementary information (Section C in Multimedia Appendix 1).

## Assessment Measures

During this phase, the assessment measures encompassed the following: time to completion data (with successful grasps serving as indicators of task completion); initial rehabilitation goals set by the physiotherapists and patients prior to commencing the 12 sessions incorporating the VR task (see Textbox 1); an evaluation of goal attainment postcompletion of the 12 sessions involving the VR task; 2 standardized motor function assessments (ie, the Box and Block Test, BBT, which measures gross manual dexterity—number of blocks moved in 60 s—and 9-Hole Peg Test, 9HPT, which assesses fine motor coordination in seconds [75]; see Table 2); and each patient's subjective experience, documented through their comments following the completion of the 12 sessions.

**Textbox 1.** Physiotherapy goals for patients A and B as reported by their physiotherapist.

**Patient A goals**

- Ability to move in an upright position
- Strengthening of leaning on the left side of the body

**Patient B goals**

- To gain confidence in walking
- Improve balance
- Muscle condition improvement

**Table 2.** Patient A and B's Box and Blocks Test and 9-Hole Peg Test scores.

| Patient | Hand | BBT[a,b] Pre (blocks) | BBT Post (blocks) | 9HPT[c,d] Pre (s) | 9 HPT Post (s) |
|---|---|---|---|---|---|
| A | Left | 27 | 27 | 21.30 | 22.40 |
| A | Right | 51 | 50 | 20.84 | 23.30 |
| B | Left | 23 | 22 | 20.77 | 21.50 |
| B | Right | 54 | 54 | 21.60 | 21.20 |

[a]BBT: Box and Blocks Test.
[b]Scores indicate the number of blocks a patient can move over a partition from one compartment to another within 60 seconds, using one hand.
[c]9HPT: Nine-Hole Peg Test.
[d]Scores provide a standardized measure of fine motor dexterity, particularly assessing hand–eye coordination, finger function, and speed of movement during a precision-based task.

Throughout these initial sessions, both physiotherapists guided their patients to begin the task by focusing their visual attention above the fireplace in the VR environment, focusing on the painting object. Once each patient's gaze was visually focused, a countdown timer, counting down from 3 to 0, initiated the appearance of a ball within the FOV. The patients were then instructed to reach out and grasp the ball as it appeared. Upon the patients' successful grasping of the ball and subsequent completion of this first trial, both patients completed an additional 14 trials before the task ended. The physiotherapists would remain present with their patients throughout the entirety of this study to provide additional support or further instructions that might be needed by the patient. This was also due to safety reasons, as both patients interacted with the task while standing upright.

## Patient Completion Time Description

We obtained 180 trials per participant. Before proceeding to data inspection, the data were trimmed to remove the worst (ie, slowest) 2 trials for each week. A visual inspection of the task completion times of both patients is presented in Figure 5. Conducting statistical analyses on 2 patients is supported within the framework of case study methodology, which allows for the test of intervention effects on an individual basis [76].

**Figure 5.** Patient task completion times. Dotted data points illustrate the completion times of each trial. Whiskers denote observed SE of the mean for each day, based on the trial-level data. The solid line signifies a loess interpolation, providing a continuous representation of the completion times over time. The dashed line indicates the trend predicted by the model. To ease visualization, dots above 20 seconds (NA=12; NB=7) are not displayed.

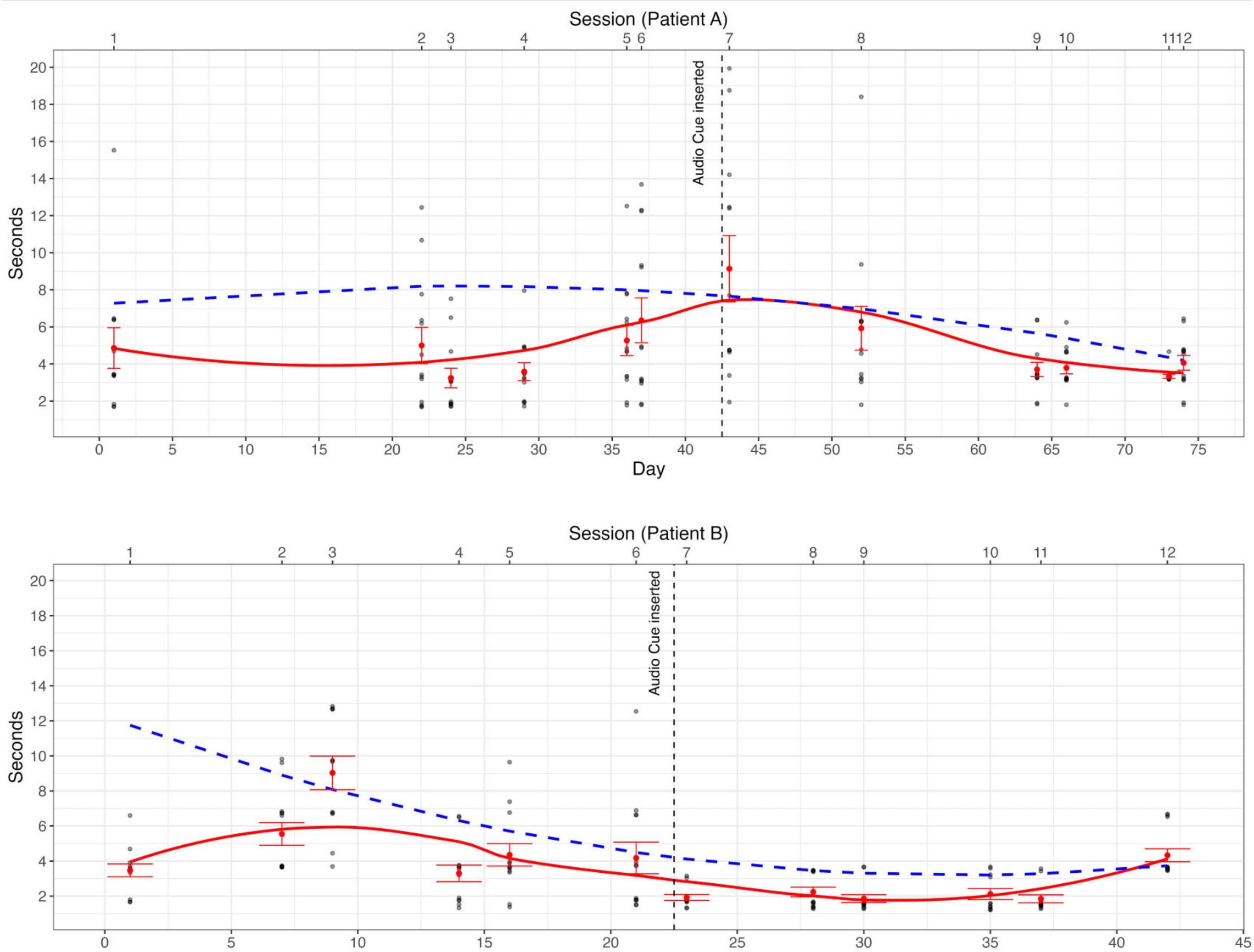

A curve estimation analysis was conducted to examine the trends of the completion times throughout the sessions (N=12 with 15 trials each) for both patients independently. Consistent with the positive skewness of the completion times (Skewness for Patient A=2.63; Skewness for Patient B=4.11), we resorted to a generalized linear model approach through R’s *lme4* package [77]. The data were modeled using a gamma distribution, which is ideal for positively skewed strictly positive continuous data [78]. To facilitate interpretation, no link function was used, allowing coefficients to be interpreted directly on the original scale of the outcome. In greater detail, 6 curve models were fitted to the data (ie, linear, logarithmic, quadratic, power, inverse growth, and exponential decay), consistent with methodologies in rehabilitation research that use curve estimation to track patient progress over time [79]. Subsequently, the models were compared based on the Akaike Information Criterion, Bayesian Information Criterion, Nagelkerke $R^2$, and performance score (through the *performance* R package [80]) (see eg, [81,82] model comparison and information criteria). In all models, the day of the session was used as the predictor. This was preferred over the session number because our sessions were not equally distant in time. For both patients, the quadratic model showed the best fit (see Table 3).

**Table 3.** Model comparison.

| Patient and model | AIC[a] | BIC[b] | $R^2$ | Performance |
|---|---|---|---|---|
| Patient A | | | | |
| linear | 1022 | 1032 | 0.070 | 50.04 |
| quadratic | 1019 | 1032 | 0.104 | 100.00 |
| inverse | 1031 | 1041 | 0.013 | 0.39 |
| log | 1029 | 1039 | 0.027 | 7.63 |
| exp | 1031 | 1041 | 0.012 | 0.00 |
| power | 1031 | 1040 | 0.016 | 1.96 |
| Patient B | | | | |
| linear | 941 | 950 | 0.146 | 47.02 |
| quadratic | 932[c] | 945[c] | 0.205[c] | 85.40[c] |
| inverse | 962 | 972 | 0.004 | 11.09 |
| log | 945 | 954 | 0.120 | 28.07 |
| exp | 962 | 972 | 0.002 | 14.87 |
| power | 957 | 966 | 0.042 | 3.50 |

[a]AIC: Akaike Information Criterion.
[b]BIC: Bayesian Information Criterion.
[c]Best models.

Regarding Patient A, when inspecting the raw data, a marked increase in completion time was found at session 7, namely, the session wherein the audio cue was added. However, the model suggested that, after a nonsignificant initial increasing trend ($\beta_{linear}$=0.08, SE=0.08; *P*=.33), a downward curvature began to emerge, approaching significance ($\beta_{quadratic}$=–0.001, SE=0.001; *P*=.08), around day 24 (ie, session number 3).

Model's results for Patient B indicate that the completion times exhibited characteristics of a significantly decreasing duration ($\beta_{linear}$=–0.53, SE=0.20; *P*=.009), where initial task completion times were followed by a gradual deceleration toward a minimum time for task completion, reached around day 35 (ie, session number 10; $\beta_{quadratic}$=0.007, SE=0.003; *P*=.04). Furthermore, Patient B also showed a strong reduction in average completion times and intraweek variability after session 6.

## Assessment of Upper Limb Motor Function

The BBT and 9HPT were included as standardized measures of gross and fine motor function, respectively, to assess upper limb performance relevant to the grasping demands of the VR task.

Table 2 summarizes pre- and postintervention scores from the BBT and 9HPT, which were performed under therapist supervision, with patients instructed to direct their focus toward their neglected side. Results show consistent right-hand performance on the BBT while the left (affected) hand showed mild impairment.

## Patients' Subjective Experience

At the conclusion of the 12-session period, Patient A and Patient B were individually asked to provide feedback regarding the use of the VR task in conjunction with their physiotherapy treatment (see Table 4 below for a summary of Patient A and B's qualitative feedback). These feedback rounds were conducted by their respective physiotherapists, each posing the same set of two questions: (1) To what extent were you able to achieve your rehabilitation goals? (2) How would you describe your experience with the VR task, using your own words?

**Table 4.** Qualitative feedback from Patients A and B in response to 2 questions posed by their physiotherapist.

| Questions | Patient A feedback | Patient B feedback |
|---|---|---|
| Were your rehabilitation goals achieved? | "In my opinion, there is no difference in my condition." | "Shifting weight to the left at the end of the period was easier and was slightly more successful. During this period, I started walking without support, without a walking stick."<br>"Walking has become more confident." |
| How did you experience the VR[a] task (in your own words)? | "The VR task was fun and interesting. The rehabilitation experience was different from my normal physiotherapy experience." | "It felt quite nice and relatively easy. When the balls were on the left *and* behind, it was more difficult. The audio did not seem to really help my performance in the game. But they calmed my thoughts down. The most challenging was when I couldn't move my legs along but, closer to the end it was easier." |

[a]VR: virtual reality.

Both physiotherapists transcribed the responses provided by their patients, with the conversations taking place in the Finnish language.

Table 4 indicates that while Patient A did not consciously notice a difference in his condition, he experienced the VR task as fun and interesting. Patient B actively noticed a difference, potentially contributed to by engagement in the VR rehabilitation task. Patient B explained that the VR technique was partially more effective and that shifting weight was easier. In addition, the patient began to walk without support following the VR tasks. Patient B also explained the difficulty in grasping for the ball when it was in the area affected by neglect yet noted that in addition to being a relatively easy exercise, the audio calmed them.

# Discussion

## Principal Findings

The objectives of this case study were 2-fold: (1) to develop a VR-based solution incorporating audiovisual cueing, designed for real-time use during physiotherapy training sessions for poststroke VSN rehabilitation and (2) to explore the initial experiences and perceptions of both patients and physiotherapists regarding the use of audiovisual cueing within this VR task during rehabilitation. The development process resulted in the successful development of a perceptual motor task customized to meet the needs of real-time physiotherapy applied adjunctly to patients with VSN. Both patients reported a positive subjective experience with the VR training, citing enjoyment and interest, and 1 patient even experienced some improvement in motor function. The VR training was also positively received by physiotherapy trainees. Taken together, these results provide several promising case-specific elements and a potential roadmap for future task development, as well as a larger trial with explicit control and standardization.

## Development Aims

The VR task incorporated a progressive approach to rehabilitation [19,20], as evident in the task design. For instance, several adjustments were made in response to user feedback (eg, accessibility considerations were integrated into the system, with a deliberate adjustment of the FOV to cater to patients with VSN). Notably, the usage of VR technology depended on the provision of training to the physiotherapists to use the technology prior to testing it with patients. This training was a vital aspect of the study, ensuring they possessed the necessary expertise to integrate technology into rehabilitation practices safely. Insights from the developer and shared experiences from physiotherapy trainees emphasize the collaborative nature of the approach, incorporating external perspectives and expertise.

## Patient A and B Qualitative Feedback

As we used a patient-centered approach, the subjective experience of patients was a key outcome measure. Generally, the qualitative feedback from both patients highlights positive engagement with the task. Both patients reported positively about their experience, stating: "the VR task was fun and interesting" (Patient A) and the VR task was "quite nice" and "relatively easy" (Patient B). The VR task offers a different experience, which was positively received, as Patient A states: "the rehabilitation experience was different from my normal physiotherapy experience." This indicates benefits beyond strictly functional and medical outcomes, such as increased patient engagement during physiotherapy training, which may find alignment with [83].

Patient B's feedback also reflects progress toward achieving rehabilitation goals (eg, the objective of regaining confidence in walking). However, Patient A perceived no significant difference in their condition, which may be due to a multitude of factors. While Patient B also notes some challenges, suggesting potential difficulties related to spatial awareness or balance training, they state that with training, the task did become easier. This may indicate that the VR task was challenging the patient with positive training outcomes as a result. The patient's comment about the game's audio—"it provides a calming effect"—while not intended, aligns with literature on sensory stimulation having a calming effect on patients [84] Though not directly impacting performance, it implies the potential for sensory engagement as a therapeutic aid. However, the functional clarity and perceived usefulness of the audiovisual cues were not explicitly evaluated in this study, pointing to the need for future iterations to include cue-specific assessment. Indeed, as indicated by Danso et al [52], more research must be done to systematically study the impact of sound and music on therapeutic progress. Despite being at distinctly different stages of poststroke recovery and presenting with varying symptomatic profiles, both patients reported a positive experience. This outcome highlights the value of customizing the VR task, including directional cues (audio and visual) and in-task settings (eg, background music volume, audio cue volume), to suit individual needs.

The continuous presence of the physiotherapists with their patient throughout the duration of the study, while using VR, suggests a patient-centric approach [24,85]. They actively assisted patients in fitting the HMD as well as manually initiating and monitoring each VR trial, providing real-time guidance and instructions to both patients. The introduction of the VR task to patients by their therapist, coupled with detailed explanations of its objectives and instructions, aligns with the literature on patient-centered education and rehabilitation practice [24]. The aspect of providing the patients with an understanding of the rehabilitation process may have contributed to both patients' positive feedback.

## Individual Differences in Task Response

While it is important to keep in mind the anecdotal nature of the evidence due to this being a case study, tentatively positive results were obtained from the patient's interaction with the VR task. The integration of task metrics, standardized motor assessments, and patient feedback highlights individual differences in response to the VR intervention.

Patient B showed a significant decay trend in task completion times, along with reduced intrasession variability and subjective reports of improved confidence in walking [17,86]. Patient A, by contrast, exhibited variable completion times and no perceived functional change. Pre- and postintervention scores on the BBT and 9HPT revealed mild left-hand impairment for both patients, with little measurable change over the 12 sessions—suggesting that motor gains alone are unlikely to account for Patient B's improved task efficiency. Notably, both patients performed within normative ranges on the 9HPT [87] for the unaffected hand, with lower scores for the affected hand, consistent with moderate upper-limb asymmetry typical of right hemisphere stroke. Furthermore, across sessions, ball bounce speed (see Table S3 in Multimedia Appendix 1, a proxy for task difficulty) was held constant, indicating that performance differences were not attributable to variation in task demands.

Although the distinct patient profiles, including comorbidities such as VSN, left-side hemiparesis, and left-side hemiplegia, may have influenced task interactions and completion times, it is critical to approach these findings with caution given the case study design. VSN involves attention and awareness deficits with perceptual components [6-8,88], while left-side hemiparesis and hemiplegia relate primarily to motor capacity, with hemiparesis indicating muscle weakness and hemiplegia signifying a complete loss of motor control. In addition, differences in the timing of intervention—Patient A, 1 year poststroke, versus Patient B, 4 years poststroke—may have contributed to variations in task outcomes. Therefore, it is essential to interpret these results as case-specific and within the limitations inherent to a case study framework.

The difference in the total VR task completion times, measured in days for the 2 patients, Patient A (74 days from Day 1 to the final session) and Patient B (42 days over the same period), tentatively suggests a shorter interval between treatment sessions is associated with faster task completion time. This finding is generally supported by the literature [21,22].

## Limitations and Future Research

The study has several significant limitations that must be considered case-specific when interpreting its results. A key constraint for the quantitative data is that these are case studies of 2 patients diagnosed with VSN. This limited participant pool, as well as limited characterization of VSN symptoms, is limited by the absence of standardized neglect assessments (eg, the Behavioral Inattention Test) and neuroimaging data (eg, magnetic resonance imaging/computed tomography lesion localization), constraining the interpretation of the underlying neural correlates of task performance in response to this VR task. Furthermore, the 2 patients were at different stages in their rehabilitation journey—one having experienced a stroke in 2018 and the other in 2021. Such disparities in their recovery timelines could introduce confounding variables, thereby impacting the generalizability of the study's findings. In addition, the task-specific outcomes for each patient might have been influenced by numerous uncontrolled variables within the study.

To address these limitations, our future research roadmap involves research including a full sample of patients with VSN (a sample size calculation will be conducted using G*Power software, Heinrich-Heine-Universität Düsseldorf [89], derived from comparable VR rehabilitation studies), as well as additional standardized assessment measures (eg, the Montreal Cognitive Assessment [90], Hospital Anxiety and Depression Scale [91], and Berg Balance Scale [92] will be used). In addition, we will incorporate neglect-specific assessments, such as the Catherine Bergego Scale [93] and Behavioral Inattention Test [94] to quantify changes in VSN. To further investigate the impact of audiovisual cueing, we also plan to incorporate a comparison condition, such as a visual-only cueing group, alongside a standard control group. This will allow us to more rigorously evaluate the specific contribution of multisensory feedback to rehabilitation outcomes.

Regarding the VR task, future analyses of task completion time at varying locations across the 30° horizontal plane will be made to compare how patient progress in reaching easier versus harder targets on the contralesional side. A subsequent study may be designed to include trials on both the left and right sides, incorporating a structured analysis to determine whether the observed improvements in task performance are due to overall speed enhancement or are specifically observed within the neglected hemispace. Although out of scope of this study, collecting richer qualitative reports from family members who regularly interact with each patient will provide clearer insights into the patients' daily activities and recovery needs [3,95], offering a holistic perspective on their progress. While this study did not focus on improvements in grasping skill, methodologies from related research [28,31] could be applied in future studies to assess such skill in patients with VSN using this VR task. In addition, using the Suite for the Assessment of Low-Level Cues on Orientation [96] could enrich our understanding of how patients with VSN perceive and navigate in VR environments, potentially offering valuable insights for rehabilitation practices. A follow-up study will include eye-tracking measurements to track visual attention during a trial, as well as include targeted evaluations of audiovisual cue clarity and perceived utility to better understand their functional role in attention orientation and motor engagement (eg, prompting gaze shifts or initiating reach movements). These improvements in study design and data collection will mitigate some of the limitations observed in this study.

Although participants did not report any discomfort or restrictions associated with the HMD equipment, future studies could explore the use of wireless HMD systems, such as the HTC Vive XR Elite, Meta Quest, or Varjo VR, to enhance patient comfort and mobility. In addition, incorporating advanced tracking technology in future studies could improve control over experimental variables.

An additional consideration for future exploration is the potential of the VR intervention to positively influence patient motivation. Both patients in our case studies responded very positively to the VR intervention, and for follow-up studies, measures of patient motivation such as the Motivation in Stroke Patients for Rehabilitation Scale [97] could be included to provide valuable insight into this aspect of the intervention. Crucially, to strengthen the evaluation of patient experience, future studies will incorporate validated instruments, such as the System Usability Scale [67] and Intrinsic Motivation Inventory [98], alongside direct patient-reported outcome measures, to reduce reliance on therapist-transcribed responses and minimize potential reporting bias. Motivation and patient enjoyment have been shown to be important drivers of positive rehabilitation outcomes [5]. The gamified nature of the VR task, coupled with the novelty of the technology, may support patient motivation, encouraging patients to persist in therapy despite the difficulty of the task.

## Conclusions

This study explored the development and implementation of a VR-based physiotherapy intervention designed for hand grasping training for VSN rehabilitation. Positive preliminary user experience reports from both patients and physiotherapists provide promising evidence for a future research roadmap of this VR task and highlight the individual patient differences in response to VR-assisted physiotherapy. In addition, the distinct responses of the 2 patients highlight the intervention's potential capacity for personalized adaptation, emphasizing its suitability for diverse VSN rehabilitation needs. To further enhance the feasibility of this VR task in physiotherapy rehabilitation, future research should focus on the use of additional standardized measures applied to a full-scale sample. To better understand the specific impact of this task on attention to the neglected hemispace, future investigations should include a structured analysis of motor performance at varying target locations across the 30° horizontal plane. Furthermore, a full-scale sample could provide further insight into whether observed improvements are due to enhanced movement speed or specific engagement with the neglected hemispace. These findings support the use of VR as a patient-centered tool that can be tailored to individual profiles, offering promising directions for future research in neurorehabilitation.

### Acknowledgments

This project has received funding from the Research Council of Finland (346210).

### Conflicts of Interest

None declared.

### Multimedia Appendix 1

Supplementary materials on virtual reality setup and sessions.
[DOCX File (Microsoft Word File), 494 KB-Multimedia Appendix 1]

## Abbreviations

**9HPT:** 9-Hole Peg Test
**BBT:** Box and Block Test
**FOV:** field of view
**HMD:** head-mounted display
**HRTF:** head-related transfer function
**MRC:** Medical Research Council
**VR:** virtual reality
**VSN:** visuospatial neglect